\documentclass[a4paper]{jpconf}
\usepackage{graphicx}
\usepackage{amssymb}
\usepackage[utf8]{inputenc}

\begin{document}
\title{The complexity of parsec-scaled dusty tori in AGN}

\author{K.~R.~W.~Tristram$^1$, M.~Schartmann$^{2,3}$, L.~Burtscher$^{4,}$\footnote[5]{Present address:
Max-Planck-Institut für extraterrestrische Physik, Gießenbachstraße, 85748 Garching,
Germany}, K.~Meisenheimer$^4$, W.~Jaffe$^6$, M.~Kishimoto$^1$, S.~F.~Hönig$^7$, G.~Weigelt$^1$}

\address{$^1$ Max-Planck-Institut für Radioastronomie, Auf dem Hügel 69, 53121 Bonn, Germany}
\address{$^2$ Max-Planck-Institut für extraterrestrische Physik, Gießenbachstr., 85748 Garching, Germany}
\address{$^3$ Universitäts-Sternwarte München, Scheinerstraße 1, 81679 München, Germany}
\address{$^4$ Max-Planck-Institut für Astronomie, Königstuhl 17, 69117 Heidelberg, Germany}
\address{$^6$ Leiden Observatory, Leiden University, Niels-Bohr-Weg 2, 2300 CA Leiden, The Netherlands}
\address{$^7$ Physics Department, University of California, Santa Barbara, CA 93106, USA}

\ead{tristram@mpifr-bonn.mpg.de}

\begin{abstract}
Warm gas and dust surround the innermost regions of active galactic nuclei (AGN). They provide the material for accretion onto the super-massive black hole and they are held responsible for the orientation-dependent obscuration of the central engine. The AGN-heated dust distributions turn out to be very compact with sizes on scales of about a parsec in the mid-infrared. Only infrared interferometry currently provides the necessary angular resolution to directly study the physical properties of this dust. Size estimates for the dust distributions derived from interferometric observations can be used to construct a size--luminosity relation for the dust distributions. The large scatter about this relation suggests significant differences between the dust tori in the individual galaxies, even for nuclei of the same class of objects and with similar luminosities. This questions the simple picture of the same dusty doughnut in all AGN. The Circinus galaxy is the closest Seyfert 2 galaxy. Because its mid-infrared emission is well resolved interferometrically, it is a prime target for detailed studies of its nuclear dust distribution. An extensive new interferometric data set was obtained for this galaxy. It shows that the dust emission comes from a very dense, disk-like structure which is surrounded by a geometrically thick, similarly warm dust distribution as well as significant amounts of warm dust within the ionisation cone.
\end{abstract}

\section{Introduction}

Molecular gas and dust play an important role in the appearance and physics of active galactic nuclei (AGN). A significant portion of the energy released by the accretion disk is intercepted by the surrounding dusty material and re-emitted at infrared wavelengths. The direct connection between the emission from the accretion disk and the reprocessed emission from the torus is reflected by the tight correlation between the X-ray and mid-infrared luminosities of AGN \cite{2008Horst1,2009Gandhi,2009Levenson}. Several lines of evidence (ionisation cones, collimated outflows, broad emission lines in polarised light, etc.) suggest that the molecular and dusty material is located in a geometrically thick, toroidal distribution, the so-called \textit{dusty torus}. The viewing-angle dependent obscuration of such a distribution -- or, in general, the supposed axisymmetric geometry of AGN -- is made responsible for many of the different observed properties of AGN, most importantly the differences between type 1 (torus face-on) and type 2 (torus edge-on) sources. There are, however, certainly other parameters apart from the viewing angle which influence the observed properties of AGN (see for example the contribution by C. Ramos Almeida).

Most information about the distribution of the nuclear gas and dust has been obtained indirectly, as even the largest single-dish telescopes do not provide enough angular resolution to resolve the nuclear material. Only in the last years it has become possible to spatially resolve the thermal emission from the AGN-heated dust using infrared interferometry. While interferometry in the near-infrared probes the hot ($T\sim1400\,\mathrm{K}$) dust close to sublimation at the inner rim of the torus (c.f.\ the contribution by M.\ Kishimoto), mid-infrared interferometry explores the warm ($T\sim300\,\mathrm{K}$) body of the dust distribution. The following will focus on studies of AGN with the MID-infrared Interferometric instrument (MIDI) at the Very Large Telescope Interferometer (VLTI). MIDI is currently the only instrument capable of resolving the emission from the warm dust component in AGN.

\section{Observations of AGN with MIDI}

Several observing campaigns of AGN have been carried out with MIDI. First, the brightest four active galaxies were studied individually, mainly using science demonstration time (SDT) or guaranteed time observations (GTO): NGC~1068 \cite{2004Jaffe1, 2009Raban}, the Circinus galaxy \cite{2007Tristram2}, Centaurus~A (NGC~5128) \cite{2007Meisenheimer} and NGC~4151 \cite{2009Burtscher}. In addition to these four sources, seven further sources were observed using the remaining GTO in a snapshot survey \cite{2009Tristram}. The goal was to test the feasibility of observations for these fainter sources at the sensitivity limit of MIDI and to obtain first size estimates of their emission regions. After the successful completion of the feasibility tests, a large observing programme containing 13 sources was initiated with the aim to investigate a statistical sample of AGN (L.~Burtscher et al.\ in prep.). A sample of six type 1 sources studied with MIDI has been presented in \cite{2011Kishimoto2}. A few other sources have been observed successfully, among them NGC~3783 \cite{2008Beckert, 2011Kishimoto2} and NGC~424 (S.~F.~Hönig et al.\ in prep.). Here, the results obtained from the snapshot survey as well as the results obtained from new measurements for the Circinus galaxy will be summarised.

A list of all the galaxies from GTO observations and the snapshot survey is shown in table~\ref{table_sources}. Both type 1 and type 2 AGN as well as a radio galaxy and a quasar are among these sources. The last column lists a short, verbal result from the interferometric observations for the respective galaxy. These results range from more or less unresolved sources (Mrk~1239) to very well resolved sources, where a disk component can be distinguished from a more extended, diffuse structure (NGC~1068 and the Circinus galaxy).

\begin{table}
\caption{\label{table_sources}List of galaxies from GTO observations and the snapshot survey successfully observed with MIDI. The columns are (1) the name, (2) the right ascension (3) the declination, (4) the type and (5) the total mid-infrared flux of the galaxy at $12\,\mathrm{\mu m}$ as well as (6)  a short, verbal summary of the results from the interferometric measurements.}
\begin{center}
\begin{tabular}{lrrlrl}
\br
Galaxy name     & RA             & DEC         & type   & $F_{\mathrm{MIR}}$ {[}Jy] & result \tabularnewline
(1)             & (2)            & (3)         & (4)    & (5)                       & (6) \tabularnewline
\mr
NGC~1068        & $02\;42\;41$ & $-00\;00\;48$ & Sy~2   & $16.5$                    & disk + elongated component \tabularnewline
NGC~1365        & $03\;33\;36$ & $-36\;08\;26$ & Sy~1.8 & $ 0.4$                    & partially resolved \tabularnewline
IRAS~05189-2524 & $05\;21\;01$ & $-25\;21\;45$ & Sy~2   & $ 0.4$                    & faint detection (resolved?) \tabularnewline
MCG-05-23-016   & $09\;47\;40$ & $-30\;56\;56$ & Sy~2   & $ 0.6$                    & well resolved \tabularnewline
Mrk~1239        & $09\;52\;19$ & $-01\;36\;44$ & Sy~1.5 & $ 0.4$                    & unresolved \tabularnewline
NGC~3783        & $11\;39\;02$ & $-37\;44\;19$ & Sy~1   & $ 0.6$                    & partially resolved \tabularnewline
NGC~4151        & $12\;10\;33$ & $+39\;24\;21$ & Sy~1.5 & $ 1.2$                    & well resolved \tabularnewline
3C~273          & $12\;29\;07$ & $+02\;03\;09$ & QSO    & $ 0.3$                    & possibly resolved \tabularnewline
Centaurus A     & $13\;25\;28$ & $-43\;01\;09$ & FR~I   & $ 1.3$                    & core + extended component \tabularnewline
IC~4329A        & $13\;49\;19$ & $-30\;18\;34$ & Sy~1.2 & $ 1.0$                    & slightly resolved \tabularnewline
Circinus        & $14\;13\;10$ & $-65\;20\;21$ & Sy~2   & $10.2$                    & disk + elongated component \tabularnewline
NGC~7469        & $23\;03\;16$ & $+08\;52\;26$ & Sy~1.2 & $ 0.6$                    & well resolved \tabularnewline
\br
\end{tabular}
\end{center}
\end{table}

\section{The size--luminosity relation in the mid-infrared}

\subsection{Building the size--luminosity relation}

From the visibilities $V$ measured by the interferometer, simple size estimates for the mid-infrared emission regions in these galaxies can be derived by assuming a Gaussian brightness distribution. Although this is certainly a very simplifying assumption and the true brightness distributions of AGN tori are significantly more complex, a Gaussian distribution is the simplest and most general initial guess for the true brightness distribution. Furthermore, it does not rely on any other assumption (e.g.\ about the location of the inner or outer edge of the torus as is needed for power-law distributions) and its half width at half maximum $\mathit{HWHM}$ is related to the visibility $V$, the projected baseline length $\mathit{BL}$ and wavelength $\lambda$ of the interferometric observation by a simple, analytical relation:
\begin{equation}
\mathit{HWHM}=\frac{\lambda}{\mathit{BL}}\cdot\frac{1}{\pi}\sqrt{-\ln2\cdot\ln V}.\label{formula_gauss-fwhm}
\end{equation}
This method yields viable size estimates with errors of up to a factor of 4 as long as $0.2 \lesssim V \lesssim 0.9$ (for a more detailed discussion of the accuracy see \cite{2011Tristram}).

\begin{figure}
\begin{center}
\includegraphics{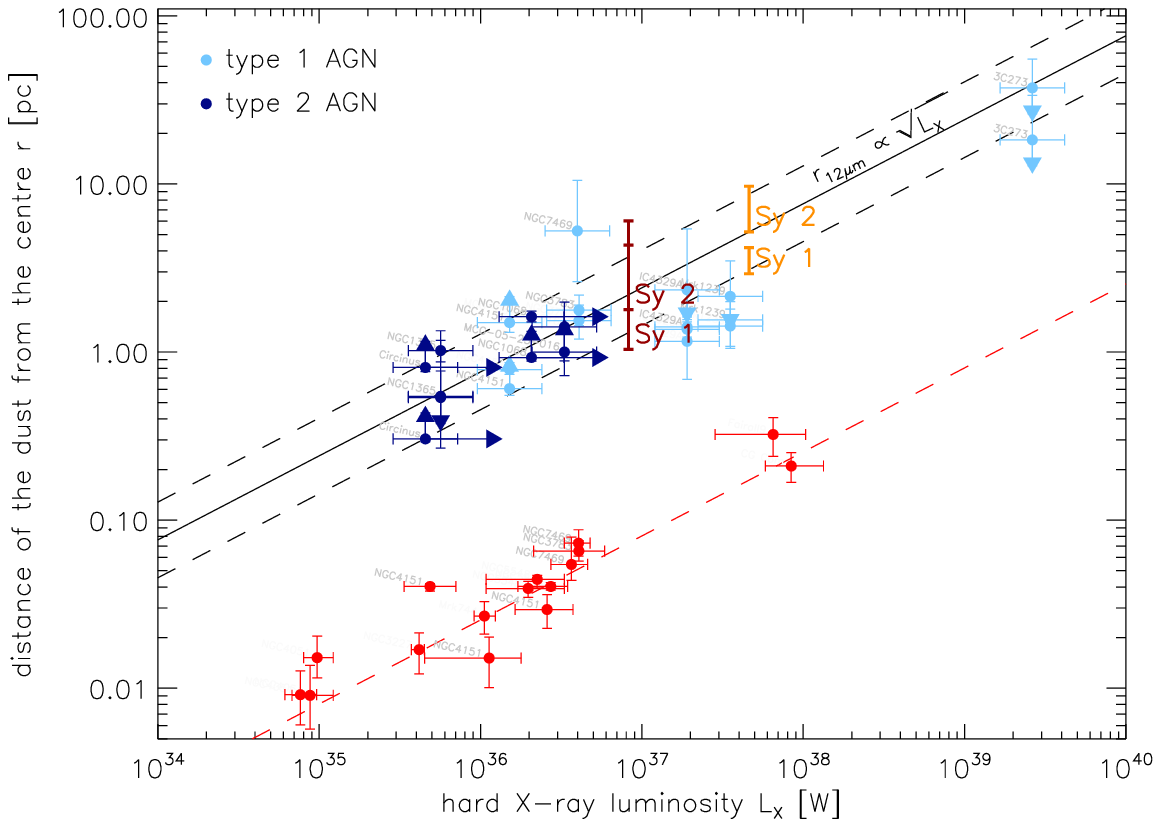}
\end{center}
\caption{\label{size-luminosity}Size of the mid-infrared emitter as a function of the X-ray luminosity, taken as a proxy for the intrinsic AGN luminosity. Type~1 objects are plotted in light blue, type~2 objects in dark blue. Upper and lower limits on the size estimates are marked by arrows. The fitted relation $r = p \cdot L^{0.5}$ is delineated by the black continuous line, the scatter in the relation by the dashed lines. The red dashed line shows the inner radius of the torus derived from reverberation measurements and near-infrared interferometry. The individual measurements of near-infrared radii are plotted as red points and were adapted from figure~30 in \cite{2006Suganuma}.}
\end{figure}

In figure~\ref{size-luminosity} the Gaussian $\mathit{HWHM}$ is plotted as a function of the hard X-ray luminosity $L_X$, which is considered as a proxy for the intrinsic luminosity of the AGN when corrected for foreground and intrinsic absorption. The X-ray fluxes were taken from the first 22 months of data of the hard X-ray survey ($14-195\,\mathrm{keV}$) with the Burst Alert Telescope (BAT) on the Swift satellite \cite{2010Tueller} and corrected for absorption. These measurements were obtained before or contemporaneous with the interferometric measurements. For the only source not in the BAT sample, Mrk~1239, the luminosity from \cite{2004Grupe} was used after correcting it from the soft to the hard X-ray band.

The data shown in figure~\ref{size-luminosity} is still consistent with the naively expected relation $r\sim L^{0.5}$, i.e.\ that the size of the dust distribution scales with the luminosity of the accretion disk: $r = (0.76\pm0.11)\cdot10^{-18}\,\mathrm{(L/W)}^{0.5}\,\mathrm{pc}$. This is in contrast to the much slower dependency than $L^{0.5}$ found for type 1 objects in \cite{2011Kishimoto2}. However, we also see a tendency for type 1 objects with high luminosities to lie below the relation and a lot depends on the highest luminosity object 3C~273. Certainly, this figure has to be filled with more measurements of a larger sample of sources and the uncertainties as well as selection effects have to be reduced. This is one of the goals of the MIDI AGN Large Programme.

\subsection{Comparison to models}

To better understand the implications of the interferometrically measured sizes of the dust distributions, the size--luminosity relation is compared to the predictions from two models for clumpy tori of AGN: (1) the clumpy torus model from \cite{2006Hoenig, 2010Hoenig2} and (2) the hydrodynamical torus model from \cite{2009Schartmann}. To obtain torus sizes which can be directly compared to the measured size estimates, interferometric observations of the brightness distributions of the two models are simulated and then the sizes are calculated according to equation~\ref{formula_gauss-fwhm}. The range of sizes predicted by the two models are shown by the thick red and orange bars in figure~\ref{size-luminosity}. The sizes from the models are in general consistent with the observed sizes at a given luminosity. The models themselves cannot be used to test whether the size--luminosity relation is correct in itself, as they are explicitly assumed to scale with $L^{0.5}$. More interesting is however that both models show a clear offset between the sizes of type 1 (more compact) and type 2 (more extended) tori. Such an offset is \textit{not} observed in this data and the dust distributions in individual galaxies of a single object class appear more different than the differences between the classes. This might be interpreted in the way that actually not so much the inclination but other torus properties are responsible for the type 1/2 differences. This is also suggested by the fitting of torus models to the infrared SEDs of AGN (see contribution by C. Ramos Almeida). In any case this is further evidence that not all AGN have the same dust distribution -- even not in nuclei with the same luminosity and in the same object class. 

\section{The dust distribution in the Circinus galaxy}

Being the second brightest AGN in the mid-infrared (c.f.\ table~\ref{table_sources}) and relatively nearby, at a distance of only $4\,\mathrm{Mpc}$, the Circinus galaxy is one of the prime targets for studying the nuclear dust distribution  in detail with infrared interferometry. Observations in the optical, X-rays, and radio indicate that the Circinus galaxy has a Seyfert 2 nucleus. As a typical type 2 AGN, the galaxy shows broad emission lines in polarised light \cite{1998Oliva} and an ionisation cone \cite{1997Veilleux,2000Wilson,2004Prieto}. The mid-infrared spectrum shows a deep silicate absorption feature, caused both by dust lanes within Circinus (which is inclined by $i=65^\circ$) as well as by the nuclear dust itself. First results obtained with MIDI have shown that the emission is resolved into a disk-like component plus an extended emission region \cite{2007Tristram2}. Since then, extensive new observations have been carried out and the first results of these are presented here. 

\begin{figure}
\begin{center}
\includegraphics{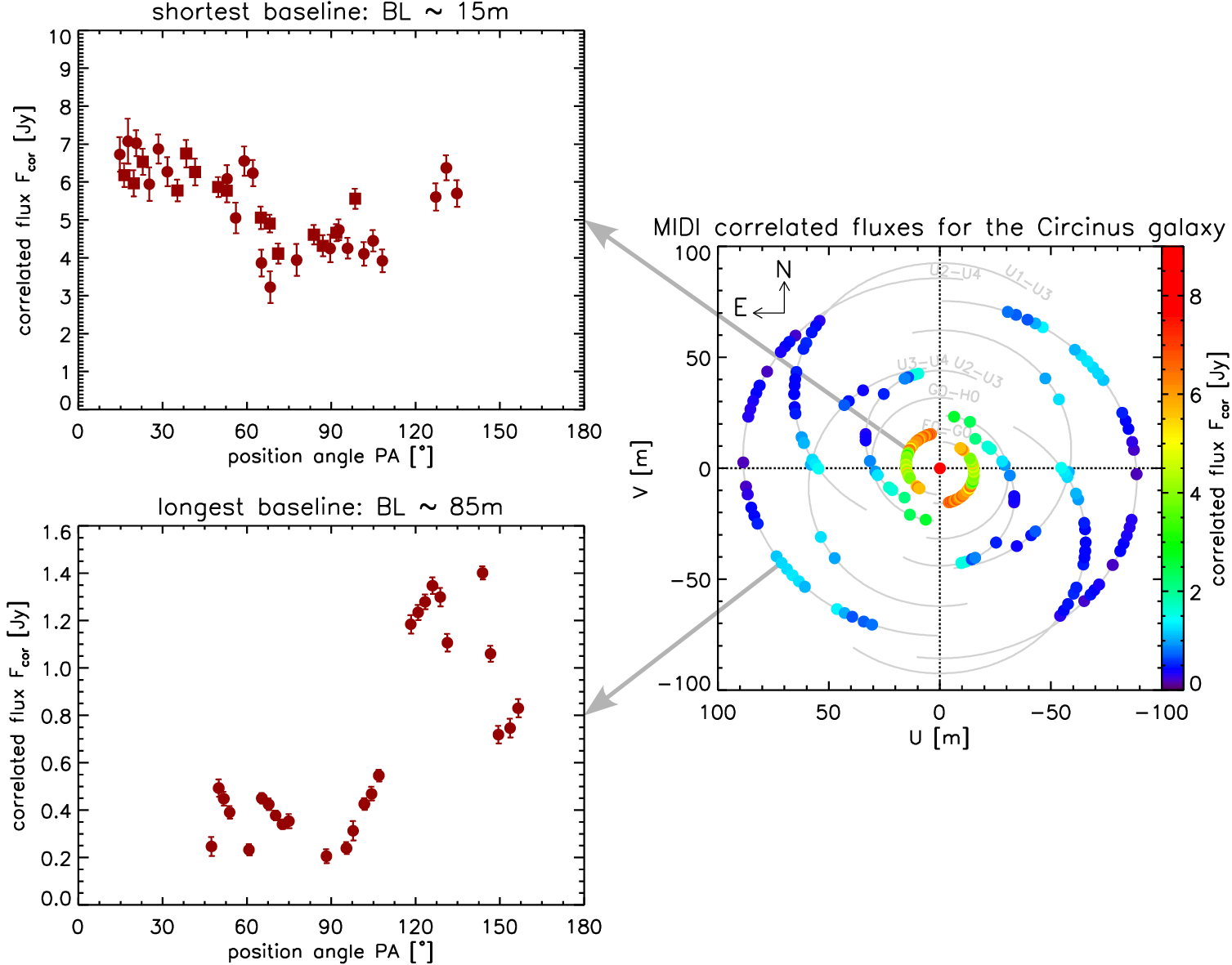}
\end{center}
\caption{\label{circinus_plots}Coverage of the UV plane for the Circinus galaxy achieved by observations with MIDI (right panel). The position in the UV plane corresponds to the angle and the length of the projected baseline of the two telescopes, when seen from the centre. The UV points are colour-coded according to their interferometrically measured correlated fluxes at $12\,\mathrm{\mu m}$. On the left hand side, these correlated fluxes are shown as a function of the position angle for two baselines with essentially constant baseline lengths (i.e.\ resolution power of the interferometer): on the top left for the shortest baseline (E0-G0), probing the largest spatial scales, and on the bottom left for the longest baseline (UT2-UT4), probing the smallest structures.}
\end{figure}

\subsection{A direct analysis of the MIDI data}

The current coverage of the UV plane achieved by MIDI is shown in the right panel of figure~\ref{circinus_plots}. With more than 100 measurements, a respectable UV coverage has been achieved for optical interferometry. However, the data are very inhomogeneous in instrument settings and observing technique, due to the large spread of time over which the data has been obtained. Furthermore, there are some inconsistencies in the data, which make a quick interpretation of the data difficult. Nevertheless, important properties of the dust emission can be directly inferred from the interferometric measurements without the need for any modelling.

On the shortest baseline, essentially only the position angle changed while the baseline length remained more or less the same. This means the same, large spatial scales were probed at different position angles. In the top left panel of figure~\ref{circinus_plots}, the correlated flux at $12\,\mathrm{\mu m}$ for this baseline is plotted as a function of the position angle. The correlated flux has a broad minimum at $\mathit{PA}\sim100^\circ$, which means the emission is resolved most along this position angle, i.e.\ it is extended most in this direction. This is consistent with the result of deep mid-infrared imaging with $8\,\textrm{m}$ class telescopes, where an elongation along $\mathit{PA}\sim100^\circ$, i.e.\ along the southern edge of the ionisation cone, is seen \cite{2005Packham}.

The same can be done for the UT2-UT4 baseline, which is a very long baseline and hence probes very small spatial structures. In the lower left panel of figure~\ref{circinus_plots}, the correlated flux from this longer baseline is plotted as a function of position angle in the same way as for the shortest baseline. A very strong dependency of the correlated flux on the position angle is evident, with a clear maximum at $\mathit{PA}\sim140^\circ$. The mid-infrared emission hence has a minimum extension along this direction. In fact, the observed dependency of the correlated flux on the position angle is consistent with a relatively thin and highly inclined disk.

\subsection{Simple two component model fit}

\begin{table}
\caption{\label{table_model-params}Best fit parameters for the two component black body Gaussian model for the dust distribution in the nucleus of the Circinus galaxy.}
\begin{center}
\begin{tabular}{lrrcrr}
\br
parameter              &\multicolumn{2}{c}{disk component}     &\hspace{1cm}&\multicolumn{2}{c}{extended component}\tabularnewline
\mr
size of major axis     & $\Delta_{1} =$   & $18\,\textrm{mas}$ &            & $\Delta_{2} =$   & $99\,\textrm{mas}$ \tabularnewline
axis ratio             & $r_{1} =$        & $0.50$             &            & $r_{2} =$        & $0.56$             \tabularnewline
position angle         & $\alpha_{1} =$   & $61^\circ$         &            & $\alpha_{2} =$   & $94^\circ$         \tabularnewline
silicate depth         & $\tau_{1} =$     & $1.4$              &            & $\tau_{2} =$     & $2.1$              \tabularnewline
temperature            & $T_{1} =$        & $295\,\textrm{K}$  &            & $T_{2} =$        & $358\,\textrm{K}$  \tabularnewline
surface filling factor & $f_{1} =$        & $1.00$             &            & $f_{2} =$        & $0.17$             \tabularnewline
\br
\end{tabular}
\end{center}
\end{table}

\begin{figure}[]
\includegraphics[width=0.5\textwidth]{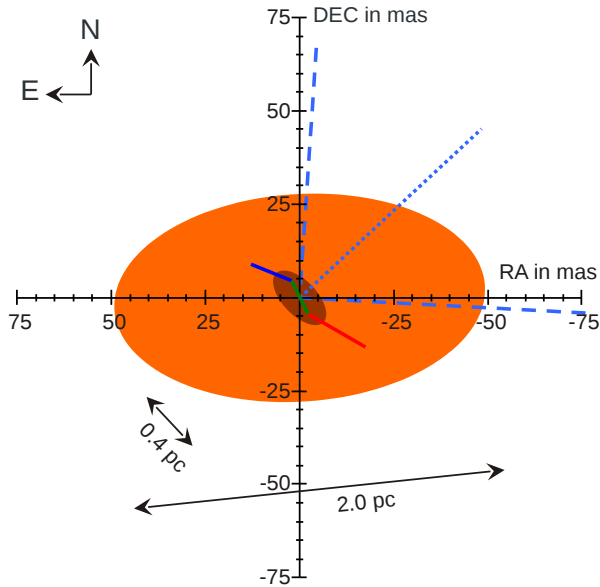}\hspace{0.05\textwidth}%
\begin{minipage}[b]{0.45\textwidth}\caption{\label{circinus_model}Sketch of simple two component model fit to the data for the Circinus galaxy: a highly elongated warm emission region with a temperature of $\sim300\,\mathrm{K}$ (brown) is surrounded by an elongated and slightly warmer emission region (orange). In the centre of the sketch, the location of the $\mathrm{H}_{2}\mathrm{O}$ maser emitters in a disk from \cite{2003Greenhill} is overplotted: the blue line to the north-east traces the receding masers and the red line to the south-west the approaching masers. The dashed blue lines trace the edge of the observed ionisation cone seen in [OIII] \cite{2000Wilson}, the dotted line is the cone axis.}
\end{minipage}
\end{figure}

To obtain a more accurate idea about the brightness distribution, the interferometric data has to be modelled. The most unambiguous result which is also physically motivated and does not rely on any further assumptions can be achieved using a two component model: The two components are elongated Gaussian distributions emitting black body or grey body radiation. Furthermore, the two components are located behind an absorption screen giving rise to the silicate absorption feature. The model thus has a total of 12 free parameters, and their values for the best fit are shown in table~\ref{table_model-params}. In figure~\ref{circinus_model}, the two model components are sketched: The inner component has a size of $20\,\textrm{mas}$, i.e.\ $0.4\,\textrm{pc}$ at the distance of Circinus. It is very elongated and has a temperature of about $300\,\textrm{K}$. The second component has a size of about $100\,\textrm{mas}$, that is $2\,\textrm{pc}$, and is elongated along $\mathit{PA}\sim94^\circ$. It has a slightly higher temperature of about $360\,\textrm{K}$. Most of the total flux of the source is located in the extended component.

We interpret the two Gaussian emitters as signs for a geometrically thick distribution, ``torus'', of warm dust surrounding a very dense, disk-like structure as well as significant amounts of dust within the ionisation cone. The smaller and denser disk is seen at a high inclination. It is optically thick and it partly exhibits the silicate band in emission. The disk matches $\mathrm{H}_{2}\mathrm{O}$ maser emitters tracing a mildly warped, thin disk \cite{2003Greenhill} in orientation and size. The disk is surrounded by the larger, less dense and most likely clumpy or filamentary torus component which gives rise to strong silicate absorption. The elongation of this component indicates that there are significant amounts of dust within the ionisation cone. Since the dust in the ionisation cone is roughly at the same temperature as the dust of the disk, this dust must be directly exposed to the radiation from the accretion disk to be heated to such high temperatures at these distances.

\subsection{Comparison with NGC~1068}

It is interesting to compare these results for the Circinus galaxy to those for NGC~1068, which hosts the brightest active nucleus in the mid-infrared and which is the second Seyfert 2 nucleus where the dust distribution has been studied in detail \cite{2004Jaffe1, 2009Raban}. There are a few similarities between the two nuclei: Both tori have a highly-inclined dust disk which is surrounded by an extended, geometrically thick dust distribution. And in both nuclei, the inner disk component roughly matches a $\mathrm{H}_{2}\mathrm{O}$ maser disk in orientation and size. However, there are also a few significant differences: While in Circinus the torus is nicely aligned perpendicular to the ionisation cone and outflow as expected in the standard picture, this is not the case in NGC~1068. While in NGC~1068 the dust disk, which is located closest to the accretion disk, has a significantly higher temperature than the surrounding dust, there are no indications for hot dust in the Circinus nucleus from our data. And finally, while in the Circinus nucleus the depth of the silicate feature decreases towards the nucleus, the opposite is the case in NGC~1068: here the silicate feature gets deeper towards the smaller disk component. Although it is still not clear what the reasons for these differences are, it is clear that the dust distributions in these two Seyfert 2 galaxies must be quite different.

\section*{Conclusions}

These proceedings should have convinced the reader of the following three points: (1) Optical interferometry is currently the only way to \textit{directly} study the nuclear dust distributions in AGN -- and it is a very successful tool to do so. (2) The warm, AGN-heated dust distributions in AGN are compact with sizes of only a few parsecs. (3) There are significant differences between the dust distributions in individual galaxies, even for objects of the same class and with similar luminosities. These differences are probably larger than the differences between the distributions in type 1 and type 2 objects. This will certainly have impact on our understanding of how accretion onto the super-massive black hole works on parsec scales.

\section*{References}
\bibliographystyle{iopart-num}
\bibliography{paper}

\end{document}